\documentclass[twocolumn]{aastex631}

\usepackage{graphicx}
\usepackage{color}

\usepackage[T1]{fontenc}
\usepackage{newtxtext}
\usepackage[varvw]{newtxmath}

\usepackage{xfrac}
\usepackage{sidecap}
\usepackage{url}
\usepackage{hyperref}
\usepackage{enumitem}

\usepackage[utf8]{inputenc}

\newlength{\figspace}
\newcommand{\beq}{\begin{equation}}
\newcommand{\eeq}{\end{equation}}
\newcommand{\beqn}{\begin{eqnarray}}
\newcommand{\eeqn}{\end{eqnarray}}

\newcommand{\St}{\mathrm{St}}
\newcommand{\Sc}{\mathrm{Sc}}

\renewcommand{\v}[1]{{\boldsymbol{#1}}} 

\renewcommand{\sun}{\odot}

\newcommand{\Eq}[1]{Eq.~(\ref{#1})}

\newcommand{\eq}[1]{\Eq{#1}}
\newcommand{\eqp}[1]{(Eq.~\ref{#1})}

\newcommand{\Fig}[1]{Fig.~\ref{#1}}
\newcommand{\Figp}[1]{(Fig.~\ref{#1})}
\renewcommand{\fig}[1]{\Fig{#1}}
\newcommand{\figp}[1]{\Figp{#1}}

\renewcommand{\table}[1]{Table~\ref{#1}}

\definecolor{brown}{rgb}{0.42,0.24,0.07}
\definecolor{darkgreen}{rgb}{0.0,0.6,0.00}
\definecolor{purple}{rgb}{0.7,0.0,0.7}
\definecolor{black}{rgb}{0.0,0.0,0.0}

\def\apjl{\rm ApJL}
\def\apjs{\rm ApJS}

\def\aap{\rm A\&A}

\submitjournal{ApJ}
\begin{document}

\title{Active Galactic Nucleus Tori: Potential Birthplace to Millions of Planets}

\author[0000-0003-0271-3429]{Bhupendra Mishra}
\affiliation{Nicolaus Copernicus Astronomical Center, Polish Academy of Sciences, Warsaw, 00716, Poland}

\author[0000-0002-3768-7542]{Wladimir Lyra}
\affiliation{New Mexico State University, Department of Astronomy, PO Box 30001 MSC 4500, Las Cruces, NM 88001, USA}

\author[0000-0002-9726-0508]{Barry McKernan}
\affiliation{Department of Science, CUNY Borough of Manhattan
 Community College, 199 Chambers Street, New York, NY 10007, USA}
 \affiliation{Department of Astrophysics, American Museum of Natural
 History, 200 Central Park West, New York, NY 10024, USA}

\author[0000-0003-0064-4060]{Mordecai-Mark Mac Low}
\affiliation{Department of Astrophysics, American Museum of Natural
 History, 200 Central Park West, New York, NY 10024, USA}

\author[0000-0002-5956-851X]{K. E. Saavik Ford}
\affiliation{Department of Science, CUNY Borough of Manhattan
 Community College, 199 Chambers Street, New York, NY 10007, USA}
 \affiliation{Department of Astrophysics, American Museum of Natural
 History, 200 Central Park West, New York, NY 10024, USA}

 \author[0000-0002-3768-7542]{Harrison E. Cook}
\affiliation{New Mexico State University, Department of Astronomy, PO Box 30001 MSC 4500, Las Cruces, NM 88001, USA}

\begin{abstract}
  The outer regions of AGN disks have temperatures similar to those of circumstellar disks, permitting dust condensation. Therefore, planet formation and growth could be active in these dust tori through similar mechanisms. We aim at quantifying the parameter space for the occurrence of streaming instability, and its outcomes in terms of the masses of the objects formed, their total number, and their continued growth via pebble accretion. We use a a recently proposed disk model with strong magnetization to keep the disk gravitationally stable. We find that the dust grain sizes required for streaming instability are easily attained through coagulation; the dust filaments it produces can contain solar masses, collapsing into tens of millions of “planetesimals” ranging from Earth to super-Jupiter masses. These planets are usually  born in the 3D Bondi regime of pebble accretion, and have mass-doubling times from 10$^3$ to 10$^7$ yrs, though 3D Hill and geometric accretion are also realized. Gas accretion occurs concurrently, and crossover mass can be attained while still in the planetary mass range. As a result, vigorous accretion can occur, leading to objects with stellar masses -- defining a core accretion channel for star formation. The pebble isolation mass is beyond the hydrogen burning limit, so accretion is limited by stellar feedback instead of gap carving. We also predict a population of exotic objects directly formed above the hydrogen burning limit, yet of pure dust. Our model suggests that AGN dust tori host the largest populations of planets in the universe.  
\end{abstract}



\section{Introduction}
\label{sec:intro}
Planets form within disks of gas and dust surrounding young stars \citep{Kant1755, Laplace1798, vonWeiszacker44, Armitage20}. However, an additional and compelling environment for planet formation has emerged \citep{Sergei12, Wada+19, Wada+21}: the disks surrounding active galactic nuclei (AGN). In recent years, significant analogies have been drawn between the physical processes within circumstellar disks and within AGN disks. These analogies suggest that mechanisms traditionally associated with planet formation may also operate in the more extreme environments of disks around supermassive black holes (SMBHs).

\citet{McKernan+12} proposed a novel scenario in which stellar-mass black holes (BHs) embedded within AGN disks can undergo hierarchical collisions and mergers, in a process analogous to the late stages of terrestrial planet formation in the core accretion model \citep{Pollack+96}---an idea grounded in the similarity of mass ratios between disk embedded objects and the central object between circumstellar and AGN disks. Just as planetary embryos accrete and migrate to mergers in disks around stars, stellar-mass BHs can accrete and migrate to mergers within gas disks orbits in the gravitational potential of an SMBH. The subsequent detection of gravitational waves from BH mergers by LIGO/Virgo \citep{Ligo15, Ligo20} including the growing number of heavyweight BH mergers within the mass-gap regime \citep{woosley17-ppi-sn, Woosley21} lends support to the formation and hierarchical merger of BHs embedded within AGN disks \citep[see e.g.][ for a review]{Ford25}.

Despite these intriguing similarities, important distinctions between AGN and circumstellar disks remain. A key difference lies in the thermal state of the disks: AGN disks are significantly hotter, leading to a higher degree of ionization due to thermal collisions. This makes them far more susceptible to the magnetorotational instability (MRI), a key mechanism for angular momentum transport in ionized disks \citep{BalbusHawley91}. In contrast, circumstellar disks are cold and weakly ionized, leading to MRI dead zones of large radial extent, and the dominant mechanism of angular momentum transport in such environments remains uncertain \citep{Lyra19, Lesur23}{,} with disk winds \citep[e.g.][]{Bai13,Gressel+15} being a leading candidate.

Another major distinction between circumstellar disks and AGN disks arises from the temperature-dependent condensation of refractories into dust grains. While circumstellar disks retain a substantial dust component throughout most of their structure except in the innermost regions where temperatures exceed the dust sublimation threshold, AGN disks are largely dust-free due to their high temperatures. 
Dust in AGN disks is confined primarily to the outer, cooler toroidal regions, collectively known as the dusty torus \citep{Antonucci85, Krolik88, Antonucci93}. This obscuring structure, located at a distance of $\sim1$--10~pc from the SMBH, plays a central role in AGN unification models and has been robustly detected via infrared interferometry \citep{Jaffe04} and subsequent radiative transfer modeling \citep{Schartmann05, Honig06, Honig17}. The torus appears to be composed of clumpy, optically thick clouds of dust and gas that reprocess short wavelength radiation from the central engine and emit strongly in the mid- to far-infrared. Observations of silicate features and spectral energy distributions provide clear evidence for dust survival in this region \citep{Hatz15}.

\defcitealias{SirkoGoodman03}{SG03}
This region poses a challenge 
for AGN disk models such as those by \citet[][hereafter \citetalias{SirkoGoodman03}]{SirkoGoodman03} or \citet{Thompson+05}. While observations confirm the presence of dust in AGN disks,  \citetalias{SirkoGoodman03}
predict temperatures across the disk that exceed the dust sublimation limit. This discrepancy arises from their assumption of a constant Toomre parameter $Q = 1$, which implies a self-regulated, gravitationally stable disk throughout. Although the exact physical mechanism that stabilizes the outer, self-gravitating regions of the AGN disk remains uncertain, one promising hypothesis involves strong magnetic pressure support
\citep{Begelman07, Jiang13, Sadowski16, Mishra20, Liska20, Mishra22, Fragile23, Tsung+25,Gerling-Dunsmore+25}.   \citet{Hopkins24} propose that magnetic fields in these regions can be significantly amplified via the advection of interstellar medium (ISM) magnetic fields, thereby providing sufficient support to prevent gravitational collapse. In our model, we adopt this framework of magnetically supported AGN disks.

The physical conditions within the dusty torus -- such as lower temperatures, enhanced densities, and the presence of shielding from the intense central radiation \citep{Elitzur06}---may allow dust grains to grow and aggregate. This creates a potentially favorable environment for planet formation, particularly at the interface between the inner AGN disk and the outer torus. In this region, where the AGN radiation field weakens and dust can persist, processes analogous to planetesimal formation in the Solar System may occur. The torus could thereby give rise to a belt of planetary mass objects, analogous to the Kuiper Belt in our own system. It is this novel idea that is the focus of this work. Here, we work out the conditions for planet formation in the dusty torus and outer disk regions of AGN.

A fundamental step in the process of planet formation is the growth of sub-micron dust grains into larger aggregates -- pebbles and planetesimals that can become gravitationally bound and collapse. 
Within the magnetically dominated, dusty, outer disk, the temperature and pressure conditions appear favorable to planetesimal formation. We investigate how the streaming instability \citep{Youdin05} can act as a trigger for the onset of planet formation \citep{Johansen+07,Bai10,Simon16} in AGN disks, and we explore the growth and long-term evolution of these bodies through pebble accretion \citep{OrmelKlahr10,LambrechtsJohansen12,Lyra+23}. Moreover, the AGN disk provides a vast reservoir of accreting gas, opening the possibility for gas accretion to fuel the growth of embedded objects, like in the classic core accretion scenario for giant planet formation \citep{Pollack+96}. Under favorable conditions, this accretion could proceed in a runaway fashion, potentially giving rise to stellar-mass objects, albeit ones formed by core accretion. These ``core accretion stars'' could continue accreting gas, perhaps even forming intermediate-mass BHs (IMBHs) via direct collapse, if accretion is not halted by stellar feedback.

This paper is organized as follows. In section \ref{sec:model} we describe the underlying equations, chosen AGN disk model, and the physics of streaming instability for planet formation. In section \ref{sec:gasaccretion}, we quantify gas accretion and its regime. In section \ref{subsec:numberofplanets}, we quantify the number of planets formed and their mass growth over the AGN lifetime. Section \ref{sec:disscussion} presents the conclusions of this study. A table of mathematical symbols used in this work is shown in \table{table:symbols}.

\begin{table*}[p]
\caption{Symbols used in this work.}
\label{table:symbols}
\begin{center}
\begin{tabular}{lll c lll}\hline
Symbol            & Definition                      & Description                     && Symbol                & Definition                          & Description \\\hline
$P_{\rm gas}$     & \eq{eq:gaspressure}             & gas pressure                    && $\xi_\mathrm{c}$               &                                     & coagulation efficiency\\
$P_{\rm mag}$     & $P_{\rm gas}/\beta_m$           & magnetic pressure               && $Z$                   &                                     & metallicity\\
$\varOmega$       & \eq{eq:keplerian-frequency}     & Keplerian frequency             && $\eta$                & \eq{eq:eta}                         & drift parameter\\
$G$               &                                 & gravitational constant          && $h$                   & $H/r$                               & disk aspect ratio\\
$M_{\rm SMBH}$    &                                 & mass of supermassive BH && $\St_{\rm SI}$        &                                     & SI threshold Stokes number\\
$r$               &                                 & distance to central mass        && $t_{\rm grow}$        & \eq{eq:growth-time}                 & coagulation timescale\\
$\rho$            & \eq{eq:volume-density}          & gas volume density              && $a_0$                 &                                     & ISM grain size\\
$T$               & \eq{eq:midplane-temperature}    & gas temperature                 && $a_{\rm SI}$          & \eq{eq:pebbleradius}                & grain size threshold for SI\\
$\dot{M}$         & $3\pi\nu\Sigma$                 & accretion rate onto SMBH        && $t_{\rm SI}$          & \eq{eq:time-to-SI}                  & time to grow from $a_0$ to $a_{\rm SI}$\\
$\sigma$          &                                 & Stefan-Boltzmann constant       && $w_\mathrm{f}$                 & \eq{eq:filament_width}              & width of filament\\
$H_{\rm ph}$      &                                 & height of disk photosphere      && $m_\mathrm{f}$                 & \eq{eq:filamentmass}                & mass of filament\\
$H$               & \eq{eq:scale-height}            & gas scale height                && $Z_\mathrm{f} $                 &                                     & metallicity of filament\\
$c_\mathrm{s} $             & \eq{eq:sound-speed}             & sound speed                     && $m_\mathrm{p}$                 & \eq{eq:mmin}                        & characteristic planetesimal mass\\
$\gamma$          &                                 & adiabatic index                 && $\delta$              &                                     & dimensionless dust diffusion\\
$k$               &                                 & Boltzmann's constant            && $R_{\rm Hill}$        & \eq{eq:Hill}                        & Hill radius\\
$\mu$             &                                 & mean molecular weight           && $R_{\rm Bondi,dust}$  & \eq{eq:RBondi-dust}                 & Bondi radius for dust\\
$m_\mathrm{H}$             &                                 & atomic mass unit                && $R_{\rm Bondi,gas}$   & \eq{eq:RBondi-gas}                  & Bondi radius for gas\\
$\varSigma$       & \eq{eq:columndensity}           & column density                  && $v_\mathrm{K}$                 & $\varOmega r$                       & Keplerian velocity\\
$T_{\rm irr}$     & \eq{eq:irradiation-temperature} & passive heating temperature     && $\Delta v$            & $\eta r$                            & Sub-Keplerian velocity reduction\\
$T_\mathrm{b}$             &                                 & background temperature          && $R_{\rm geo}$         & \eq{eq:geo}                         & geometric radius\\
$r_\star$         &                                 & disk flux effective radius      && $M_\mathrm{t}$                 & \eq{eq:Mt}                          & transition mass\\
$T_\star$         &                                 & disk flux effective temperature && $M_{\rm HB}$          & \eq{eq:Mthill}                      & Hill-Bondi transition mass\\
{$t_\mathrm{B}$}       &{$=R_{\rm Bondi,dust}/\Delta v$}& {Bondi time}             && $M_{\rm BG}$          & \eq{eq:Mtgeo}                       & Bondi-geometric transition mass\\
{$\hat{R}$}   & \eq{eq:rhat}                    & {low-St accretion radius}   && $H_\mathrm{d}$                 & $H\sqrt{\frac{\delta}{\St+\delta}}$ & dust scale height\\
{$\psi$}      & \eq{eq:psi}                     & {exponential cutoff}        && $R_{\rm acc}$         & \eq{eq:racc}                        & accretion radius\\
$\St$             & \eq{eq:stokes}                  & Stokes number                   && $\xi$                 & \eq{eq:3d2dtransition}              & 2D-3D transition parameter\\
$t_\mathrm{s}$             &                                 & stopping time                   && $t_\mathrm{p}$                 & \eq{eq:passingtime}                 & passing timescale\\
$a$               &                                 & grain radius                    && $\dot{m}_Z$           & \eq{eq:pebbleaccretionrate}         & dust mass accretion rate\\
$\rho_\bullet$    &                                 & internal grain density          && $\dot{m}_{XY}$        & \eq{eq:gasaccretion}                & gas mass accretion rate\\
$\St_{\rm frag}$  & \eq{eq:stfrag}                  & fragmentation Stokes number     && $\rho_{\mathrm{d}0}$           &                                     & dust density at midplane\\
$\St_{\rm drift}$ & \eq{eq:stdrift}                 & drift Stokes number             && $\delta v$            & $\Delta v+\varOmega R_{\rm acc}$    & approach velocity\\
$\alpha$          & $\nu/(c_\mathrm{s} H)$                    & dimensionless viscosity         && $M_{\rm thermal}$     & $3 h^3 M_{\rm SMBH}$                & thermal mass\\
$v_{\rm frag}$    &                                 & fragmentation threshold speed   && $\varSigma_{\rm gap}$ & \eq{eq:sigmagap}                    & column density at gap\\
$K$               & \eq{eq:k-gap}                   & gap density reduction parameter && $\Sc$                 & $\alpha/\delta$                     & Schmidt number\\
$\Pi$             & \eq{eq:bigpi}                   & drift parameter                 && $\tilde{G}$           & \eq{eq:Gtilde}                      & dimensionless gravity parameter\\\hline
\end{tabular}
\end{center}
\end{table*}

\section{Magnetized Outer Disk Model}
\label{sec:model}
We use 
the magnetized AGN disk model of \citet{Hopkins24} as a base to investigate planet formation in AGN disks, modified to allow for the formation of dust.
This model differs from 
the SG03 model, which assumes that the Toomre $Q$ parameter remains unity in outer regions and hence the sound speed $c_\mathrm{s} $ become constant in this marginally gravitationally unstable region. \citet{Hopkins24} instead assume that instead of radiation pressure maintaining gravitational stability, the outer region is stabilized by a strong toroidal magnetic field with $\beta_\mathrm{m} \ll 1$
, where $\beta_\mathrm{m} \equiv P_\mathrm{gas}/P_\mathrm{mag}$ and $P_\mathrm{gas}$ and $P_\mathrm{mag}$ are gas and magnetic pressure, respectively. The gas model is given by
\begin{eqnarray}
\varOmega &=& \sqrt{\frac{GM_{\rm SMBH}}{r^3}}\label{eq:keplerian-frequency}\\
\rho  &=& \frac{\varOmega^2}{2\pi G}\label{eq:volume-density}\\
T^4 &=&  T_{\rm irr}^4 + T_\mathrm{b}^4 \label{eq:midplane-temperature}\\ \label{eq:accretion-temperature}
T_{\rm irr}^4 &=& T^4_\star\left(\frac{r_\star}{r}\right)^2
\left\{{\rm max}\left[\frac{2}{3\pi}\frac{r_\star}{r} + \frac{H_{\rm ph}}{r}\left(\frac{d\ln H}{d\ln r}-1\right)\right],0\right\}\nonumber\\
&&\label{eq:irradiation-temperature}\\
c_\mathrm{s}  &=& \sqrt{\frac{\gamma kT}{\mu m_\mathrm{H}}}\label{eq:sound-speed}\\
H &=& \frac{c_\mathrm{s} }{\varOmega}\sqrt{1+\beta_m^{-1}} \label{eq:scale-height}\\
\varSigma &=& 2H\rho\label{eq:columndensity}\\
P_{\rm gas}&=&\rho c_\mathrm{s} ^2/\gamma\label{eq:gaspressure}
\end{eqnarray}
The symbols in these equations are given in Table \ref{table:symbols}.

The temperature model in \eq{eq:midplane-temperature} is equivalent to the one employed in \citet{Alarcon+24}, except without the radial flux. {W}e include here passive heating of the outer disk by {irradiation from} the hot inner disk, {with} flux given by \eq{eq:irradiation-temperature} \citep[see e.g.][]{Fukue13,Alarcon+24}. We assume this flux emanates from a characteristic radius $r_\star$ with associated effective temperature $T_\star$; by itself, this flux would heat up the torus to a temperature $T_{\rm irr}$. In this model, we assume $r_\star$ to be the innermost stable circular orbit (ISCO) and use the effective temperature at ISCO to define $T_\star$. The first term in \eq{eq:irradiation-temperature} is the grazing angle, associated with the finite size of the source, whereas the second term is the flaring angle, associated with the flaring of the disk. Taking the maximum prevents the flux from going negative, which would amount to cooling. $H_{\rm ph}$ is the height of the disk photosphere, which we take to be $4H$ \citep{ChiangGoldreich97}.  The flaring angle $d\ln H/d\ln r$ is dynamic, but we simplify it to 3/2, so in practice the irradiation flux is never negative. Finally $T_\mathrm{b}$ in \eq{eq:midplane-temperature} is the temperature of a background thermal bath. We exclude active viscous heating as it is not expected to dominate the energy budget in the dusty torus. Conveniently, without viscous heating, the model does not need to explicitly consider opacity, which is implicit in the height $H_{\rm ph}$ assumed for the disk photosphere.

\begin{figure*}
    \centering
    \includegraphics[width=\textwidth]{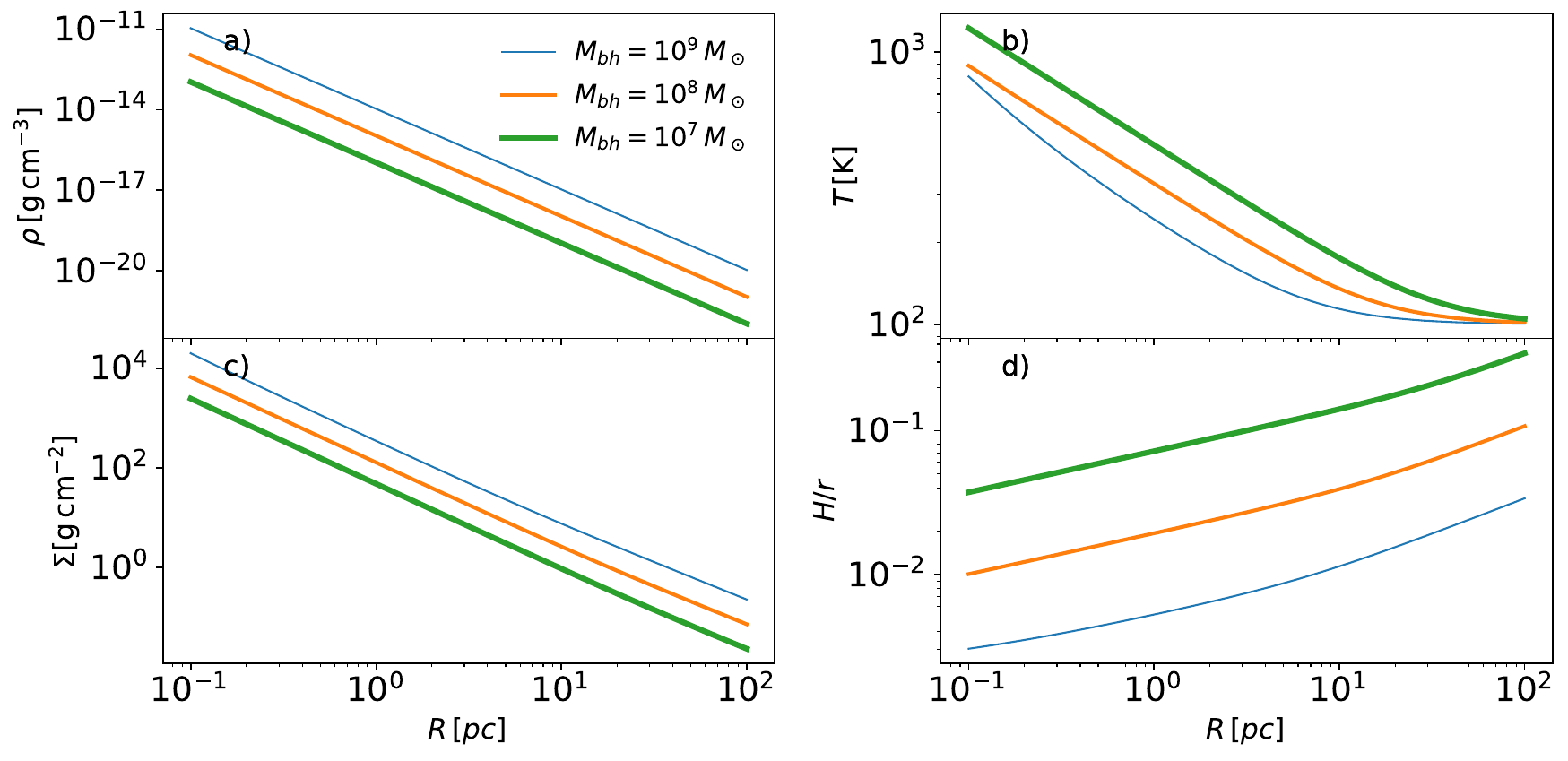}
    \caption{Radial disk profiles of relevant quantities for an AGN disk with three different central SMBH masses $M_{\rm SMBH} = 10^9,\,10^8$, and $10^7\,M_\odot$ (following the legend), and magnetic plasma $\beta_m = 0.01$. The viscosity parameter is fixed to $\alpha = 10^{-4}$ to define the mass accretion rate; the hydrogen abundance in the disk is $X=0.7$. The top row shows density ($\rho$) and temperature ($T$). The second row shows surface density ($\varSigma$) and disk aspect ratio ($H/r$).}    \label{fig:sg03mbh}
\end{figure*}

Crucially, unlike the SG03 model, this model allows the temperature to decrease in the torus, to a regime that allows for dust formation. In order to maintain a gravitationally stable disk, the model assumes a strongly magnetized flow \citep{Begelman07, Hopkins24} with $\beta_m = 0.01$. The magnetic pressure support leads to a thick disk \citep{Zhu18,Mishra20} with a scale height profile given by \eq{eq:scale-height}. Radial profiles of relevant quantities are shown in \fig{fig:sg03mbh}.

\subsection{Streaming instability}
\label{subsec:streaminginstability}
The dusty gas in a strongly magnetized AGN disk could be subject to streaming instability, which will mechanically concentrate the dust grains. This can in turn lead to the critical number density for gravitational collapse of the ensemble of grains, and thus planetesimal formation. In this section, we {quantify} the conditions for streaming instability. 

\subsubsection{Maximum Stokes number and dust grain size}
\label{subsubsec:streaminginstability}
The conditions for the streaming instability are usually defined in terms of the stopping time, the timescale for the dust to couple to the gas
\begin{equation}
  t_\mathrm{s} = \frac{a \rho_\bullet}{c_\mathrm{s}   \rho},
\label{eq:stopping}
\end{equation}
\noindent where grain size and solid density are $a$ and $\rho_\bullet$, respectively. \eq{eq:stopping} is Epstein drag of subsonic free molecular flow \citep{Epstein1924}. It assumes that the mean free path of the {molecules} is larger than the grain size, that {they} follow a thermal Maxwell-Boltzmann distribution, and that the relative motion between grains and {molecules} is subsonic. Equivalently, we define the Stokes number, a nondimensionalization of the stopping time in terms of the orbital time (or Keplerian frequency),
\begin{equation}
  \St \equiv \varOmega t_\mathrm{s}
\label{eq:stokes}
\end{equation}

Dust growth in the outer regions of the disk will be limited by either fragmentation or drift \citep{Birnstiel+12}, leading to maximum Stokes numbers given by, respectively
\begin{eqnarray}
  {\rm St_{frag}} &=& \frac{v^2_\mathrm{frag}}{3\delta c^2_\mathrm{s}},  \label{eq:stfrag}\\
 {\rm St_{drift}} &=& \frac{3\pi^{1/2} \xi_\mathrm{c} Z}{4\eta},\label{eq:stdrift}
\end{eqnarray}
where $\delta = 10^{-4}$ is the dimensionless diffusion parameter, analogous to the definition of $\alpha$ as a dimensionless viscosity parameter. Their ratio is the Schmidt number $\Sc\equiv\alpha/\delta$. Further, $v_\mathrm{frag} = 1$\,m\,s$^{-1}$ is the assumed fragmentation speed limit, $\xi_\mathrm{c}$ is the coagulation efficiency, $Z$ is the metallicity, and
\begin{equation}
\eta \equiv \frac{h^2}{2}  \left\vert \frac{\partial \ln P}{\partial \ln r } \right\vert
\label{eq:eta}
\end{equation}
is a dimensionless quantity parameterizing the radial pressure gradient in the disk and thus the drift velocity of dust. Here, $h\equiv H/r$ is the disk aspect ratio, and $P=P_{\rm mag}+P_{\rm gas} = \left(1+\beta_m^{-1}\right)P_{\rm gas}$ is the total pressure. Alternatively one can also define the parameter

\beq
\Pi \equiv \frac{\eta}{h}.
\label{eq:bigpi}
\eeq

Notice that this formulation is equivalent to $\Pi \equiv \eta v_\mathrm{K}/c_\mathrm{s}  \times 1/\sqrt{1+\beta_m^{-1}}$, i.e., scaled by $\sim\sqrt{\beta_m}$ as compared to the usual non-magnetic definition as the Mach number of the sub-Keplerian drift. For narrative clarity, we present the dust results as we derive them from the model. Fig.\ \ref{fig:streaminginstabilitymbhStSI1e-2vfrag100}a shows the drift- and fragmentation-limited Stokes number for the models considered. The grains are fragmentation-limited, and generally remain 
below the $\St_{\rm SI}=10^{-2}$ threshold (black dot-dashed line) for streaming instability favored by current numerical simulations \citep[e.g.][]{Carrera+15,LiYoudin21,Lim+24,Lim+25,Lim+26}, except in the very outer disk, beyond 10\,pc. Fig.\ \ref{fig:streaminginstabilitymbhStSI1e-2vfrag100}b shows the physical grain radius corresponding to the derived St. Grains will enter the AGN {as ISM grains, from a few nanometers to sub-$\mu$m size}, and grow to the size indicated by the fragmentation barrier.

\subsection{Dust coagulation time}

\begin{figure*}
    \centering
    \includegraphics[width=\textwidth]{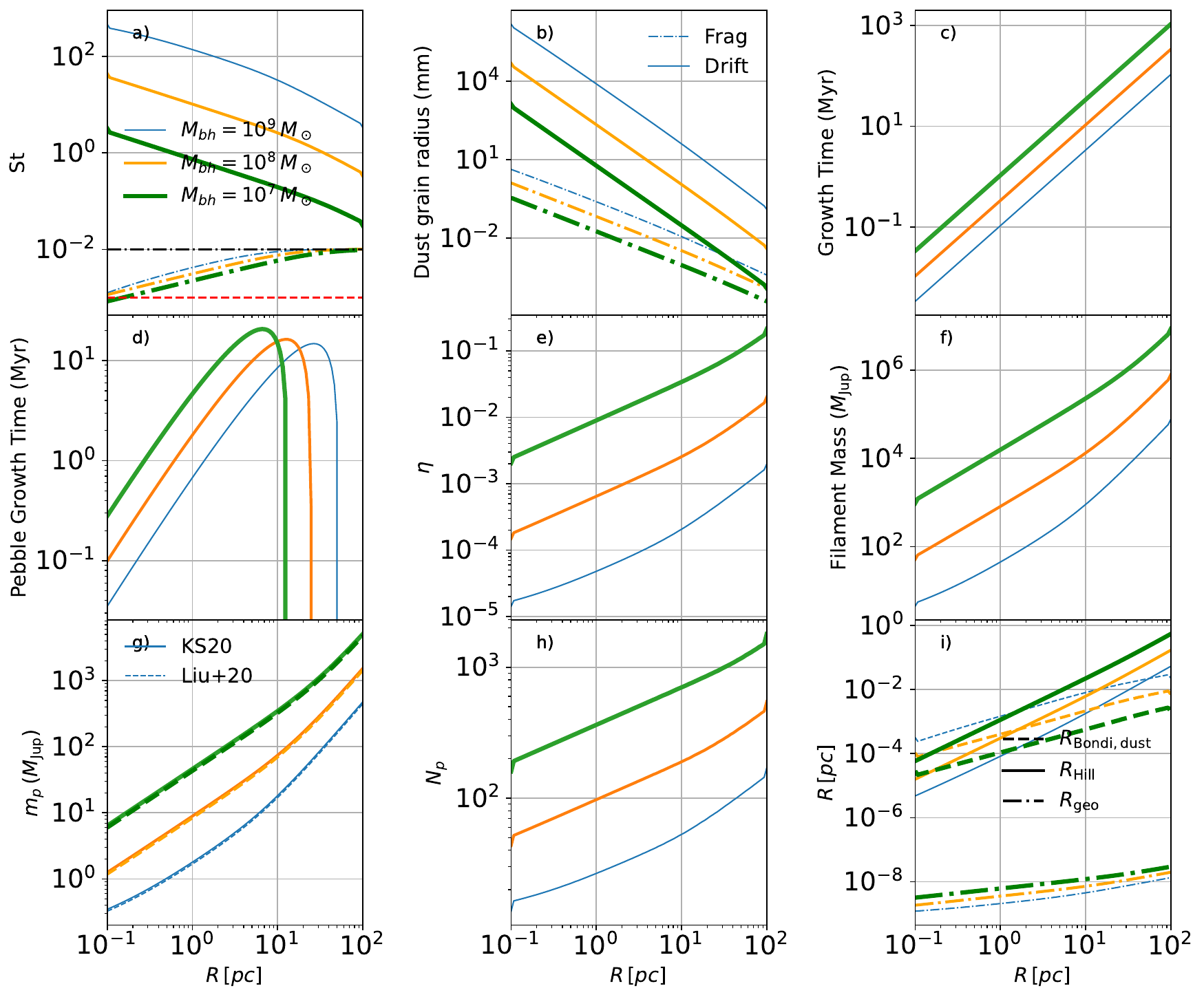}
      \caption{{Streaming instability parameters based on the AGN disk profile shown in \fig{fig:sg03mbh} and $v_\mathrm{frag} = 1\,\mathrm{m\,s^{-1} }$. The central SMBH mass varies as $M_{\rm SMBH} = 10^9\,,10^8$ and $10^7\,M_\odot$ for a fixed $\alpha=10^{-4}$ (the thickness and color of the curves show the different BH masses). The top row shows Stokes number St and dust grain radius limits due to drift {\em (solid)} and fragmentation {\em (dash-dot)}, and the coagulation time \eqp{eq:growth-time}. The second row shows pebble growth time (Eq.~\ref{eq:pebbleradius} with $\St_{\rm SI} =10^{-2}$),  the sub-Keplerian reduction parameter $\eta$ \eqp{eq:eta}, and filament mass \eqp{eq:filamentmass}. The last row shows characteristic planetesimal mass \eqp{eq:mmin} and \eqp{eq:mminLiu} (solid curves for \citealt{KS20} and dashed curves for \citealt{Liu+20}), the total number of planetary-mass objects per filament $N_p$, and the three different radii of influence in the context of accretion. The dashed red and dot-dashed black lines in the Stokes number panel (top left) represent $\St_{\rm SI} =10^{-3}$ and $\St_{\rm SI} =10^{-2}$, respectively, for a reference.}}
    \label{fig:streaminginstabilitymbhStSI1e-2vfrag100}
\end{figure*}

\begin{figure*}
    \centering
    \includegraphics[width=\textwidth]{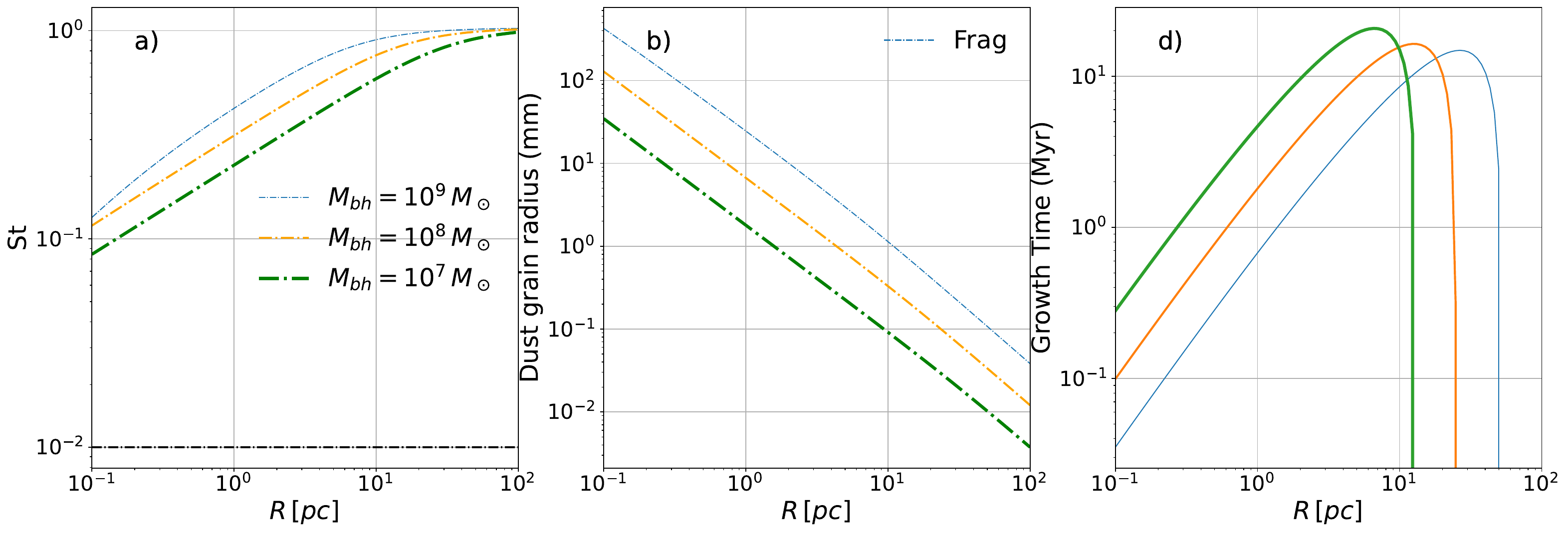}
    \includegraphics[width=\textwidth]{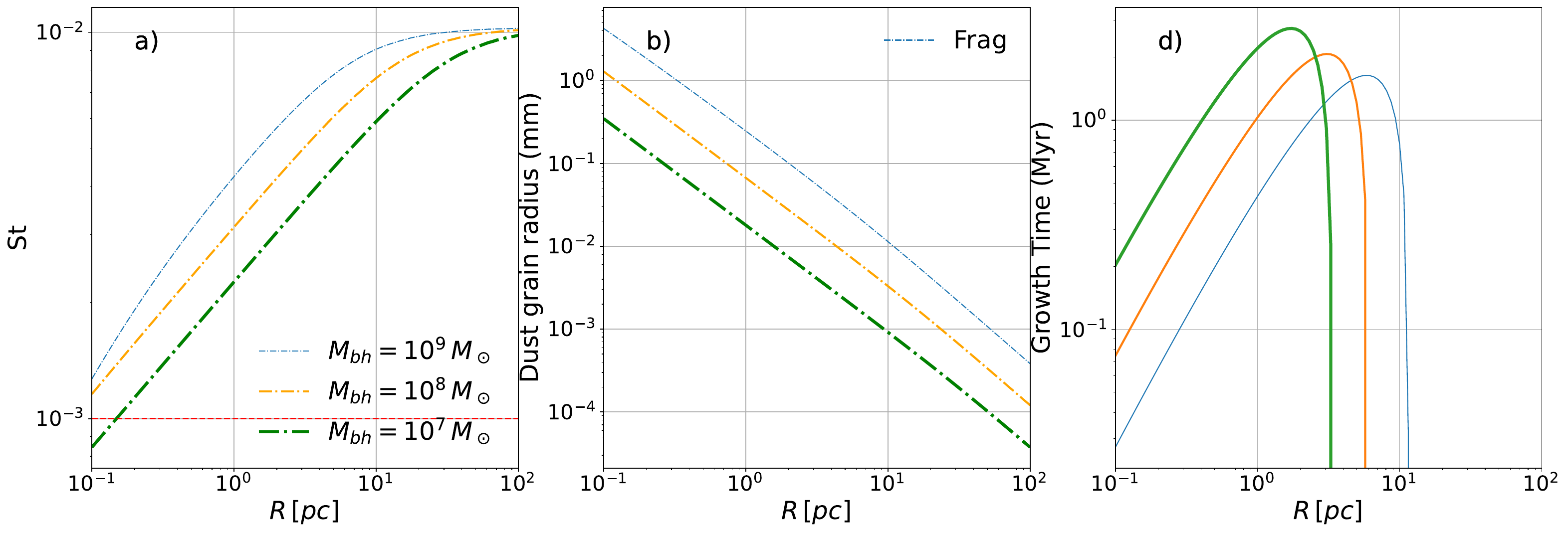}
      \caption{{Similar to Fig.~\ref{fig:streaminginstabilitymbhStSI1e-2vfrag100}a,  \ref{fig:streaminginstabilitymbhStSI1e-2vfrag100}b and \ref{fig:streaminginstabilitymbhStSI1e-2vfrag100}d but for $v_\mathrm{frag} = 10\,\mathrm{m\,s^{-1} }$ (top panel) and $v_\mathrm{frag} = 1\,\mathrm{m\,s^{-1} }$ and $\St_{\rm SI} =10^{-3}$ (bottom panel).}}
    \label{fig:streaminginstabilitymbhStSI1e-2vfrag1000}
\end{figure*}

How long does it take to grow from sub-micron ISM dust to the Stokes numbers required for streaming instability? To answer that, we calculate the coagulation e-folding time $t_{\rm grow}\equiv a/\dot{a}$ of the dust \citep{Lambrechts+14}
\begin{equation}
  t_{\rm grow} = \frac{2}{\pi^{1/2} \xi_\mathrm{c} Z \varOmega},  
\label{eq:growth-time}
\end{equation}
which yields a strong 3/2 power-law dependence on radius (Fig.\ \ref{fig:streaminginstabilitymbhStSI1e-2vfrag100}c). We consider a coagulation efficiency \citep{Brauer+07} of $\xi_\mathrm{c}=0.5$, and calculate the time to grow to pebbles at the Stokes number threshold for streaming instability, $\St_{\rm SI}$
\begin{equation}
t_{\rm SI} = t_{\rm grow}  \log \left(a_{\rm SI}/a_0\right),
\label{eq:time-to-SI}
\end{equation}
where $a_{\rm SI}$ is a pebble radius defined using $\St_{\rm SI}$ as
\begin{equation}
    a_{\rm SI} = \St_{\rm SI} \frac{c_\mathrm{s}  \rho}{\varOmega\rho_\bullet}.
    \label{eq:pebbleradius}
  \end{equation}

This is shown in Fig.\ref{fig:streaminginstabilitymbhStSI1e-2vfrag100}d, assuming $a_0=1\,{\mu}$m. The times in this figure should be contrasted with the AGN lifetime to decide whether ISM grains can grow to the size needed for streaming instability. The drop in the figure occurs when $a_{\rm SI} \leq a_0$. Depending on the chosen value of $v_\mathrm{frag}$ (see Fig.\ \ref{fig:streaminginstabilitymbhStSI1e-2vfrag1000}), the fragmentation limit Stokes number could be above or below $\St_{\rm SI}$ and allow planetesimal formation for above the threshold Stokes number ($\St_{\rm SI}=10^{-2}$). As we mentioned before, the fiducial case of $\St_{\rm SI}=10^{-2}$ with $v_\mathrm{frag} = 1\,\, \mathrm{m\, s^{-1}}$ gives a fragmentation Stokes number which is usually below the threshold value, making it difficult to form planets, unless in the very outskirts of the disk.

Increasing the fragmentation speed to $v_\mathrm{frag} = 10\, \mathrm{m\, s^{-1}}$, the model becomes bimodal, in the sense that grain growth is fragmentation-limited in the inner disk but becomes drift-limited in the outer disk, as usual in circumstellar disks. These models allow for streaming instability across the disk, yet the timescales for dust growth may be prohibitive, depending on AGN lifetime, approaching 10\,Myr around 10\,pc. However, lowering the streaming instability threshold to $\St_{\rm SI}=10^{-3}$ and keeping $v_\mathrm{frag} = 1\, \mathrm{m\, s^{-1}}$ allows the grains to again exceed the size threshold. This model too allows for planetesimal formation throughout.

\subsection{Mass of streaming instability filaments}

The saturated state of the streaming instability is characterized by
overdense filaments that can gravitationally collapse into planetesimals if dense enough.
Having established the radii in the disk where dust can grow to the
streaming instability threshold within the lifetime of the disk, we next calculate the mass of the resulting filaments.

To calculate the mass of a filament, we estimate its area and column density. A filament will have width 
\begin{equation}
    w_\mathrm{f} = \eta r
    \label{eq:filament_width}
\end{equation}
and an azimuthal length
of about a disk scale height $H$ \citep{Schaefer24}. The mass of the filament is then
\begin{equation}
m_\mathrm{f} = \eta r H \varSigma Z_\mathrm{f} ,  
\label{eq:filamentmass}
\end{equation}
\noindent where $\eta$ is defined by \eq{eq:eta} and $Z_\mathrm{f} $ is the dust-to-gas ratio of the filament (which can be different than $Z$). The $\eta$ parameter is shown in Fig.\ \ref{fig:streaminginstabilitymbhStSI1e-2vfrag100}e, and the filament mass is shown in Fig.\ \ref{fig:streaminginstabilitymbhStSI1e-2vfrag100}f. In the outer disk, a single filament can contain {millions (tens of thousands)} of Jupiter masses of pure dust for a $10^7\,M_\odot$ ($10^9\,M_\odot$) {central mass}.

\subsection{Characteristic planetesimal mass}

 The dust in the filament will collapse gravitationally and produce planets if it reaches Roche density 
\begin{equation}
    \rho_{\text{Roche}} \equiv \frac{9\varOmega^2}{4 \pi G}.
    \label{eq:rochedensity}
\end{equation}
The filament mass will not produce a single planet, but will instead fragment into many objects, with a characteristic size distribution.
The characteristic planetesimal mass formed in a simulation is given by Eq.~12 of \citet{Liu+20}, a scaling derived from the free parameters used
in streaming instability simulations
\beq
m_\mathrm{p} = C \left(\frac{Z_\mathrm{f} }{0.02}\right)^{0.5} \left(\frac{\tilde{G}}{\pi^{-1}}\right)^{0.5} \left(\frac{\Pi}{0.05}\right)^3 \hat{M}.
\label{eq:mp-liu-code}
\eeq
In the equation above, $C$ is a dimensionless coefficient,
\beq
\tilde{G}\equiv 4\pi G \rho/\varOmega^2
\label{eq:Gtilde}
\eeq
is the dimensionless gravity parameter, and
\beq
\hat{M}\equiv \rho H^3
\label{eq:hatm}
\eeq
is a mass unit. These simulations are scale-free, and the planetesimal masses are measured in units of $\hat{M}$. Thus, the masses formed should scale seamlessly from protoplanetary disks to AGN. Isolating $\rho$ in \eq{eq:Gtilde}, substituting it in \eq{eq:hatm} and plugging into \eq{eq:mp-liu-code} we find the planetesimal mass scaled to AGN parameters
\begin{eqnarray}
  m_\mathrm{p} &=& 3.1 M_\mathrm{Jup}\left(\frac{C}{10^{-5}}\right)\left(\frac{Z_\mathrm{f} }{0.02}\right)^{1/2}\left(\frac{4\pi^2 G \rho}{\varOmega^2}\right)^{3/2} \nonumber\\
  &&\times \left( \frac{h}{0.05}\right)^3 \left(\frac{M_{\rm SMBH}}{10^8 M_\odot}\right).
 \label{eq:mminLiu}
\end{eqnarray}
\noindent Here we use the dimensionless coefficient $C=1.82\times 10^{-5}$ taken from the highest resolution run of Table~2 of \citet{Schafer+17}.  \eq{eq:mminLiu} says we should expect the order of magnitude of the mass of the objects analogous to planetesimals produced by streaming instability in AGN disks to be of order that of Jupiter. This characteristic mass is shown in Fig.\ \ref{fig:streaminginstabilitymbhStSI1e-2vfrag100}g as the dashed lines. A different expression is provided by \citet{KS20} -- arguing that planetesimal formation should be regulated by diffusion, they calculate a characteristic mass
\begin{equation}
m_\mathrm{p} = \frac{1}{9}  \left(\frac{\delta}{\rm St}\right)^{1.5}
h^3  M_{\rm SMBH}, 
 \label{eq:mmin}
\end{equation}

\noindent which has the same scaling with aspect ratio and central mass as  \eq{eq:mminLiu}. The masses given by \eq{eq:mmin} are shown in Fig.\ \ref{fig:streaminginstabilitymbhStSI1e-2vfrag100}g as the solid lines. As seen in the figure, the masses found with the two estimates are similar.

We note that a minimum mass would be more informative if we are interested in the number of planetesimals formed. However, that value is unfortunately not informed by simulations, since planetesimal formation simulations notoriously do not converge at the low mass range, with more objects of increasingly lower mass forming as the resolution is increased \citep{Johansen+11,Simon+16,Lyra+24,Schafer+24}. In the remaining analysis, we shall use {as planetesimal mass} the masses calculated using \eq{eq:mminLiu}. Fig.~\ref{fig:streaminginstabilitymbhStSI1e-2vfrag100}h shows the number $N_p = m_\mathrm{f}/m_\mathrm{p}$ of planetesimals formed per filament.






\subsection{Dust accretion}

Once a planetesimal is formed, it will start scooping up the pebbles not used for planetesimal formation \citep{Lyra+08b,Lyra+09a,Lyra+09b}, a process known as pebble accretion \citep{OrmelKlahr10,LambrechtsJohansen12,Lyra+23}. Gas accretion occurs concurrently, but pebble accretion can lead to substantial growth before gas accretion begins in earnest. As such, we first check how far the planetesimals grow by pebble accretion before turning to gas accretion. 

\subsubsection{Radii of influence}

Typically there will be three different radii of influence for an embedded gravitating object whose order can vary. 
First is the Hill radius, within which material is bound to the planet against the tides from the SMBH
\begin{equation} \label{eq:Hill}
    R_\mathrm{Hill} = r \left( \frac{m_\mathrm{p}}{3M_\mathrm{SMBH}}\right)^{1/3}.
\end{equation}
Second is the Bondi radius, within which material is bound to the planet against its own kinetic energy. Since it depends on the velocity of the material, there is a Bondi radius for the pebbles, which are moving with respect to the planet with velocity $\Delta v \equiv \eta v_\mathrm{K}$, yielding
\begin{equation} \label{eq:RBondi-dust}
    R_\mathrm{Bondi, dust} = \frac{2 G m_\mathrm{p}}{\left(\eta v_\mathrm{K}\right) ^2}
 \end{equation}
and a Bondi radius for the gas. The molecules move with bulk velocity $\Delta v$ but also have their own thermal kinetic energy, moving near the sound speed. Typically, $c_\mathrm{s}  \gg \eta v_\mathrm{K}$ (or equivalently, $\beta_m/(1+\beta_m)\gg h^2$, which is satisfied for our choice of parameters), so
 \begin{equation} \label{eq:RBondi-gas}
R_\mathrm{Bondi,gas} = \frac{2 G m_\mathrm{p}}{c^2_\mathrm{s}}.
  \end{equation}
The third radius is the geometric radius
\begin{equation} \label{eq:geo}
    R_\mathrm{geo} = \left( \frac{3 m_\mathrm{p}}{4\pi \rho_\bullet}\right)^{1/3}.
\end{equation}
These radii are plotted in Fig.\ \ref{fig:streaminginstabilitymbhStSI1e-2vfrag100}i. We see they are ordered $R_{\rm geo} < R_{\rm Bondi,dust},R_{\rm Hill}$, so geometric accretion can in general be ignored. Usually, the smallest of the latter two, Bondi or Hill, defines the accretion regime. For the $10^7\,M_\sun$ central mass, $R_{\rm Bondi,dust} < R_{\rm Hill}$ throughout the entire domain. This indicates that pebble accretion will likely occur in the Bondi regime (
although this depends on the distribution of Stokes numbers of the dust grains). The disk around the other central masses will have mixed regimes of Hill and Bondi accretion.

The transition from Bondi accretion to Hill accretion occurs at \citep{Ormel17}
  \beq
  M_{\rm HB} = \frac{M_\mathrm{t}}{8 \St},
  \label{eq:Mthill}
  \eeq
  and the transition from geometric to Bondi accretion occurs at
    \beq
    M_{\rm BG} = \frac{M_\mathrm{t}}{8} \St,
    \label{eq:Mtgeo}
  \eeq
  where
  \beq
  M_\mathrm{t}=\frac{\Delta v^3}{G\varOmega}  = \eta^3 M_{\rm SMBH}.
  \label{eq:Mt}
  \eeq

  In Fig.~\ref{fig:mdot_hillbondiSt1e-2Vfrag100} we show the planetesimal mass $m_\mathrm{p}$ and the transition masses $M_{\rm HB}$ and $M_{\rm BG}$, for the three considered central masses. The Stokes number considered here is the fragmentation Stokes number, given we are in the fragmentation-limited regime. The accretion radius is \citep{Ormel17,JohansenLambrechts17}

  \beq
  R_{\rm acc}=\hat{R} e^{-\psi}
  \label{eq:racc}
  \eeq

\noindent {with the radius $\hat{R}$ depending on the accretion regime} 

\beq
\hat{R} = \left\{
\begin{array}{ll}
\left(\frac{\St}{0.1}\right)^{1/3} R_{\rm Hill} & {\rm if} \ m_\mathrm{p} \geq M_{\rm HB},\\ 
\left(\frac{4 t_\mathrm{s}}{t_\mathrm{B}}\right)^{1/2} R_{\rm Bondi,dust} & {\rm if} \ m_\mathrm{p} < M_{\rm HB},
\end{array} 
\right.  
\label{eq:rhat}
\eeq

\noindent {where $t_\mathrm{B}=R_{\rm Bondi,dust}/\Delta v$ is the time to cross the Bondi radius. Here, the argument of the exponential cutoff is} 

\beq
\psi \equiv -c_1\left(\frac{t_\mathrm{s}}{t_\mathrm{p}}\right)^{c_2}
\label{eq:psi}
\eeq

\noindent {where $c_1= 0.4$ and $c_2= 0.65$ are coefficients determined empirically via numerical simulations. Finally,}

  \beq
  t_\mathrm{p} = \frac{Gm_\mathrm{p}}{\left(\Delta v + \varOmega R_{\rm Hill}\right)^3}
  \label{eq:passingtime}
  \eeq

\noindent {is the characteristic timescale for the interaction between the pebble and the planetesimal.}

  

  The other scale of interest is $H_\mathrm{d}=H\sqrt{\delta/(\St+\delta)}$, the scale height of the pebble layer. If the accretion radius $R_{\rm acc}$ is much smaller than the pebble scale height, then the accretion regime is 3D, as the whole circular cross section is offered for accretion. In the opposite limit, of $R_{\rm acc}$ much bigger than the scale height, the accretion regime is 2D, as the planet accretes the whole column density, with the  diameter of the accretion radius as 1D cross section. \citet{Lyra+23} computed the quantity governing the transition as

  \beq
  \xi\equiv \left(\frac{R_{\rm acc}}{2H_\mathrm{d}}\right)^2.
  \label{eq:3d2dtransition}
  \eeq

%




Since $t_\mathrm{p}$ depends on the mass of the planetesimal, the equation $R_{\rm acc} - 2H_\mathrm{d} = 0$ has to be solved numerically to find the transition mass $m_{\rm 2D}$ between 3D and 2D accretion. We show this mass in the panels of Fig.~\ref{fig:mdot_hillbondiSt1e-2Vfrag100}, as solid lines, again for $\St=\St_{\rm frag}$. 

{We see from \fig{fig:mdot_hillbondiSt1e-2Vfrag100} that the accretion for the case of $v_{\rm frag}=1$\,m/s is deep in the 3D regime for all central mass cases. The accretion regime for the $10^9\,M_\sun$ case is Hill in the inner disk, and Bondi in the outer disk. For the $10^8\,M_\sun$ case the accretion regime is Bondi throughout, and for the $10^7\,M_\sun$ it is Bondi in the inner disk and geometric in the outer disk. The 3D Hill regime is a curious regime that is rarely realized in protoplanetary disks, where the transition masses $M_\mathrm{t}$ and  $m_{\rm 2D}$ are often similar.
}

{The 2D-3D transition mass $m_{\rm 2D}$ depends very sensitively on the choice of fragmentation velocity threshold $v_\mathrm{frag}$, due to its effect on the Stokes number. As we increase $v_\mathrm{frag}$ from $1 \, \mathrm{m\, s^{-1}}$ to $10\, \mathrm{m\, s^{-1}}$ (second row in Fig.\ \ref{fig:mdot_hillbondiSt1e-2Vfrag100}), the transition mass $m_{2D}$ drops by five orders of magnitude, $M_\mathrm{HB}$ also drops by about two orders of magnitude, and $M_\mathrm{BG}$ increases by two orders of magnitude.} 



 {For the case of $v_\mathrm{frag}$ from $10 \, \mathrm{m\, s^{-1}}$, the accretion mode straddles both 3D and 2D at the Hill regime for the $10^9\,M_\sun$ case; Hill, Bondi, and geometric near the 3D/2D boundary for the $10^8\,M_\sun$ case, and most 3D geometric for the $10^7\,M_\sun$ case, except for the very inner disk, that accretes at Bondi.} Since we cannot distinguish between 2D and 3D at our level of approximation, we use the general monodisperse solution of \cite{Lyra+23} for the pebble accretion rate
  \beq
  \dot{m}_Z = \pi R^2_{\rm acc} \ \rho_{\mathrm{d}0} \ \delta v \ e^{-\xi} \ \left[I_0(\xi) + I_1(\xi)\right], 
  \label{eq:pebbleaccretionrate}
  \eeq

\noindent {where $\delta v=\Delta v + \Omega R_{\rm acc}$.} The Hill radius and disk height also set the thermal mass (or rather, magnetic mass since now $h$ is mostly set by magnetic pressure) $M_\mathrm{thermal} \equiv 3 h^3 M_\mathrm{SMBH}$. If $m_\mathrm{p} < M_\mathrm{thermal}$, the length scales order as $R_\mathrm{Bondi,gas} < R_\mathrm{Hill} < H$. 

The objects are formed deeply embedded and, except for those in the geometric regime, undergo pebble accretion. The pebble accretion rates are shown in Fig.~\ref{fig:dustacc-timescale}. They hover around 10$^{-6} \ M_{\rm jup}$/yr for the {Bondi accretion rate for $v_\mathrm{frag}= 1\,\mathrm{m\,s^{-1}}$ and 10$^{-3} \ M_{\rm jup}$/yr} for $v_\mathrm{frag}= 10\,\mathrm{m\,s^{-1}}$. The right panel shows the accretion timescale $m/\dot{m}_Z$ for  the formation mass. Growth is expected for timescales comparable to the lifetime of the disk {for both cases of fragmentation speeds}. Assuming a lifetime around $10^7$ yr, all but the objects formed in the very outskirts of the disk at several tens of parsecs are expected to grow significantly by pebble accretion.
  
\subsubsection{Stellar feedback, not pebble isolation mass}

Pebble accretion continues until the planet carves a partial gap in the disk, generating a pressure gradient reversal that traps the pebbles, impeding their accretion. Simulations show that this happens when the ratio of column density in the gap compared to the unperturbed disk is $\approx$0.85. The ratio is empirically found to be \citep{Fung+14}

\beq
\frac{\varSigma_{\rm gap}}{\varSigma} \equiv \frac{1}{1+0.04 K}
\label{eq:sigmagap}
\eeq

\noindent where
\beq
K \equiv \left(\frac{m_\mathrm{p}}{M_{\rm SMBH}}\right)^2 h^{-5} \alpha^{-1}
\label{eq:k-gap}
\eeq

\noindent and $\varSigma_{\rm gap}/\varSigma=0.85$ means $K=75/17 \approx 4.4$. We solve this equation to find the planetary mass at the pebble isolation mass, shown as dot-dashed lines in the panels of Fig.\ \ref{fig:mdot_hillbondiSt1e-2Vfrag100}. Curiously, the pebble isolation mass is higher than the hydrogen burning limit (shown as grey dotted line) in all cases. That means that in the AGN case, pebble accretion will be halted not by gap carving, but by stellar feedback.


\section{Gas accretion}
\label{sec:gasaccretion}

Next we check if concurrent gas accretion happens while the objects are accreting pebbles, following \citet{Pollack+96}. Starting from the mass continuity equation, if a mass $dm$ is added, the radius increases by $dr$

\begin{equation}
dm = 4\pi r^2 \rho dr
\label{eq:continuity}
\end{equation}

\noindent dividing \eq{eq:continuity} by $dt$

\begin{equation}
\frac{dm}{dt} = 4\pi r^2 \rho \frac{dr}{dt}
\end{equation}

\noindent if $r=R_{\rm Bondi,gas}=Gm_\mathrm{p}/c_\mathrm{s} ^2$, the Bondi radius for gas accretion, then $dr/dt = G \dot{m}/c_\mathrm{s} ^2 $
so

\begin{eqnarray}
\dot{m}_{XY} &=& 4\pi R^2_{\rm Bondi,gas} \ \rho G \dot{m}_{z}/c_\mathrm{s} ^2  \nonumber \\
&=& 4\pi G^3 m_\mathrm{p}^2 \rho \dot{m}_z/c_\mathrm{s} ^6  \label{eq:gasaccretion}
\end{eqnarray}

\noindent i.e., the pebble accretion rate and gas accretion rate scale by a factor $4\pi G^3 m_\mathrm{p}^2 \rho /c_\mathrm{s} ^6$.

This should be valid as long as $R_{\rm Bondi,gas} < R_{\rm Hill}$, and the temperature at the gas Bondi radius is the same as in the nebula. Although approximate, this formula matches the six order of magnitude difference between solid and gas accretion rate in \citet{Pollack+96}. Conversely, in the case $R_{\rm Hill} < R_{\rm Bondi,gas}$, the expression would become instead
\beq
\dot{m}_{XY} = \frac{4\pi \rho G}{9\varOmega^2} \dot{m}_z
\eeq
\noindent which matches the expression with the gas Bondi radius at the thermal mass. We plot \eq{eq:gasaccretion} in the same plot as the pebble accretion rate in \fig{fig:dustacc-timescale}. As seen, the gas accretion rates are 3--4 orders of magnitude lower than the pebble accretion rate. The ratio between gas and pebble accretion rates is $4\pi G^3 m_\mathrm{p}^2 \rho /c_\mathrm{s} ^6$, which scales with the square of the planet mass. This means that the planets would have to grow in mass by a factor $\sim 30$--100 for gas accretion and pebble accretion to be comparable. Only beyond this will gas accretion outpace pebble accretion. 



\section{How many planetesimals form in AGN disks?}
\label{subsec:numberofplanets}

The total number of planetary mass objects in an AGN disk depends on the time required for dust coagulation to trigger the streaming instability, the filament masses from the nonlinear stage of the streaming, and the pebble accretion time onto the resulting objects formed by gravitational instability of the filaments. Fig.\ \ref{fig:streaminginstabilitymbhStSI1e-2vfrag100}c shows the typical range for the coagulation time ranges from a few thousand years to a few hundred million years depending on radius and central BH mass.

We can quantify the total number of massive planetesimals in an AGN disk using the number of filaments and the number of objects formed per filament. The width of filaments $w_\mathrm{f}$ is given by \eq{eq:filament_width}.
Using our numerical radial grid used for computing AGN disk profile, we estimate the number of filaments per radial bin \figp{fig:cumsum_totalplanets}. The total number of planetesimals per filament is defined by the ratio $m_\mathrm{f} / m_\mathrm{p}$, which we computed using Equations~(\ref{eq:filamentmass}) and~(\ref{eq:mmin}). Using the number of filaments per radial bin and number of planetesimals per filament, we computed the total number of planetesimals. In \fig{fig:cumsum_totalplanets}, we quantify the cumulative sum over radius of the number of filaments and planetesimals. We estimate that of order tens of millions of planetary mass objects can form in outer regions of AGN accretion disks.

\begin{figure*}
    \centering
    \includegraphics[width=\textwidth]{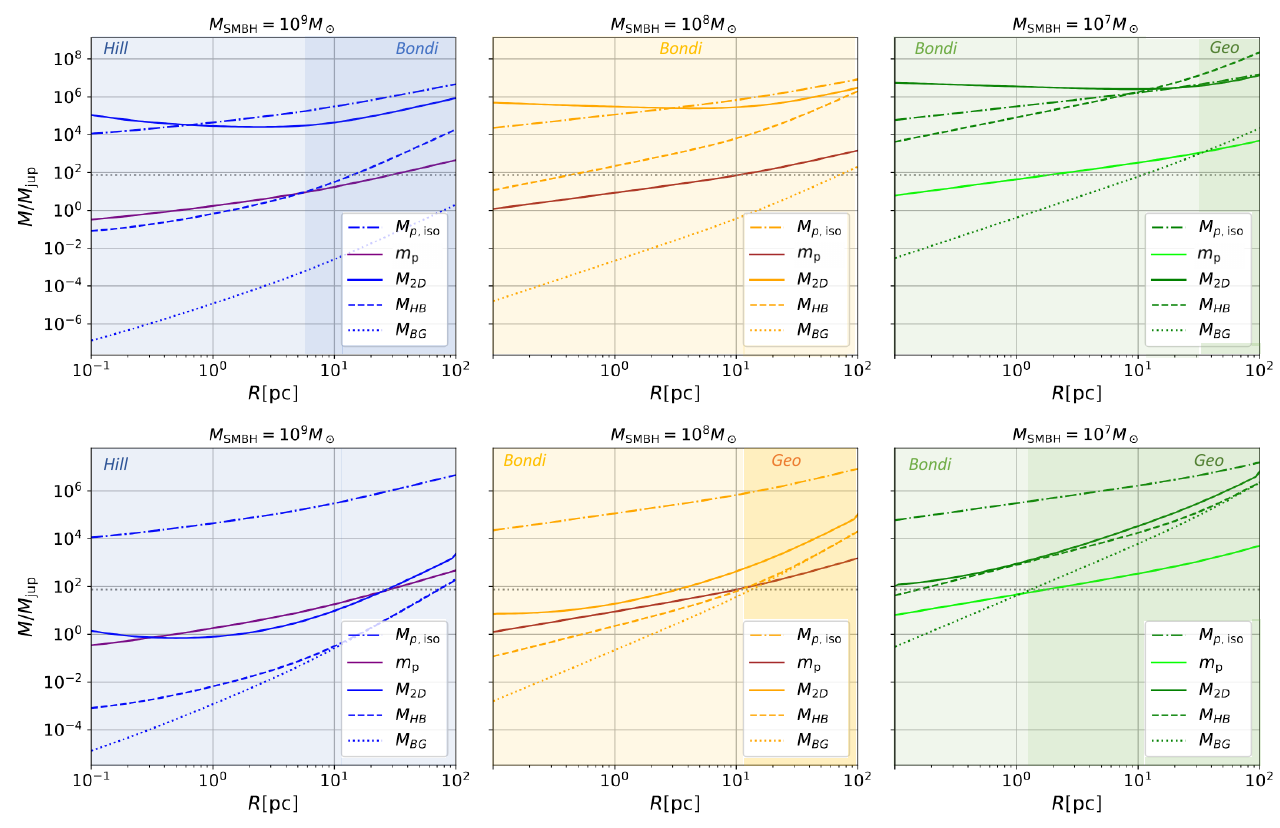}
    \caption{Transition between different accretion regimes as shown by the shading and top labels. From left to right the panels show central AGN masses of $10^9$, $10^8$ and $10^7\, M_\odot$. The top row shows $v_\mathrm{frag} = 1\, \mathrm{m\,s^{-1} }$ and the bottom row shows $v_\mathrm{frag} = 10\, \mathrm{m\,s^{-1} }$. The symbol $m_\mathrm{p}$  gives the characteristic planetesimal mass, $M_\mathrm{HB}$ is the transition mass between the Hill and the Bondi accretion regime, $M_\mathrm{BG}$ is the transition mass between the Bondi and the geometric accretion regimes, and $M_\mathrm{p, iso}$ is the pebble isolation mass. The transition between 2D and 3D accretion is also shown. In general, the objects will accrete in the 3D regime right after formation, except in the 1--10 pc range in the case $v_{\rm frag}=10$\,m/s, $M_{\rm SMBH}=10^9\,M_\sun$ (lower left). Accretion is also more efficient in the inner disk than in the outer disk, with a progression from Hill, Hill/Bondi, Bondi, and Bondi/Geometric as the central mass object decreases. The horizontal dotted black line is the hydrogen burning limit.}
     \label{fig:mdot_hillbondiSt1e-2Vfrag100}
\end{figure*}

\begin{figure*}
  \centering
      \includegraphics[width=\columnwidth]{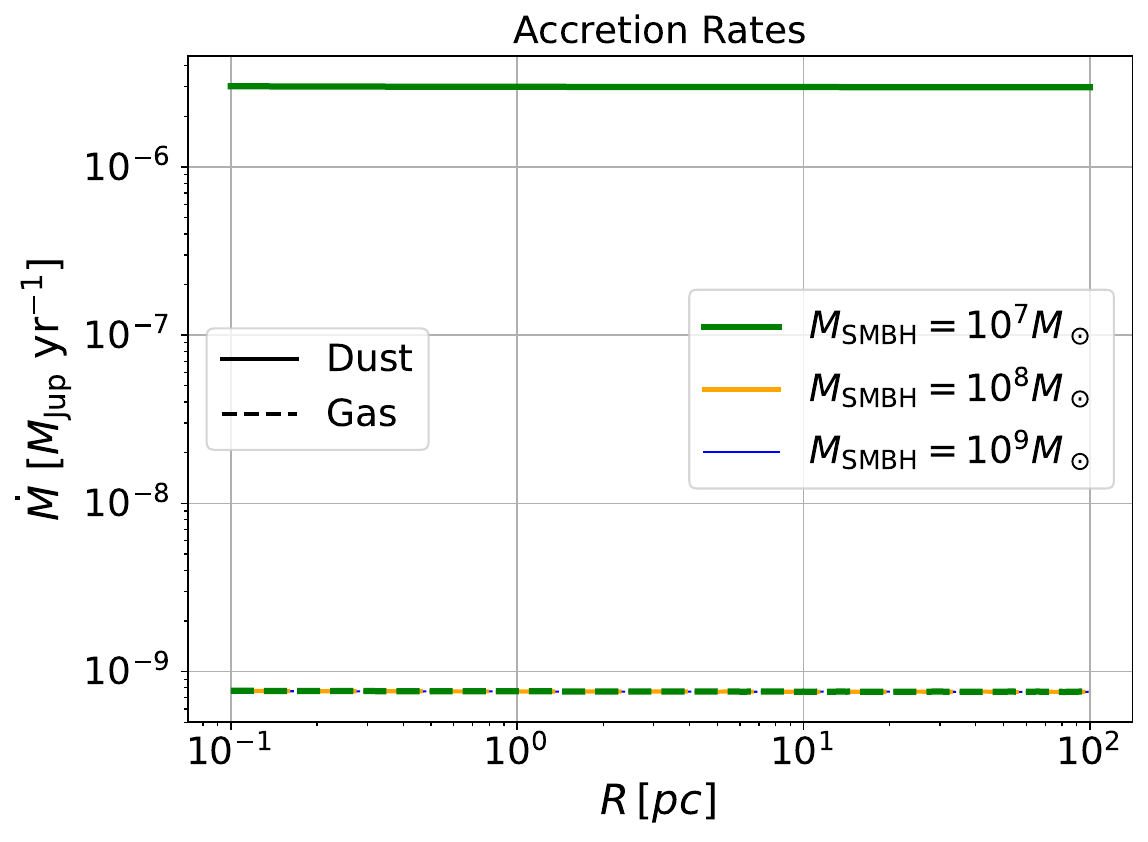}
      \includegraphics[width=\columnwidth]{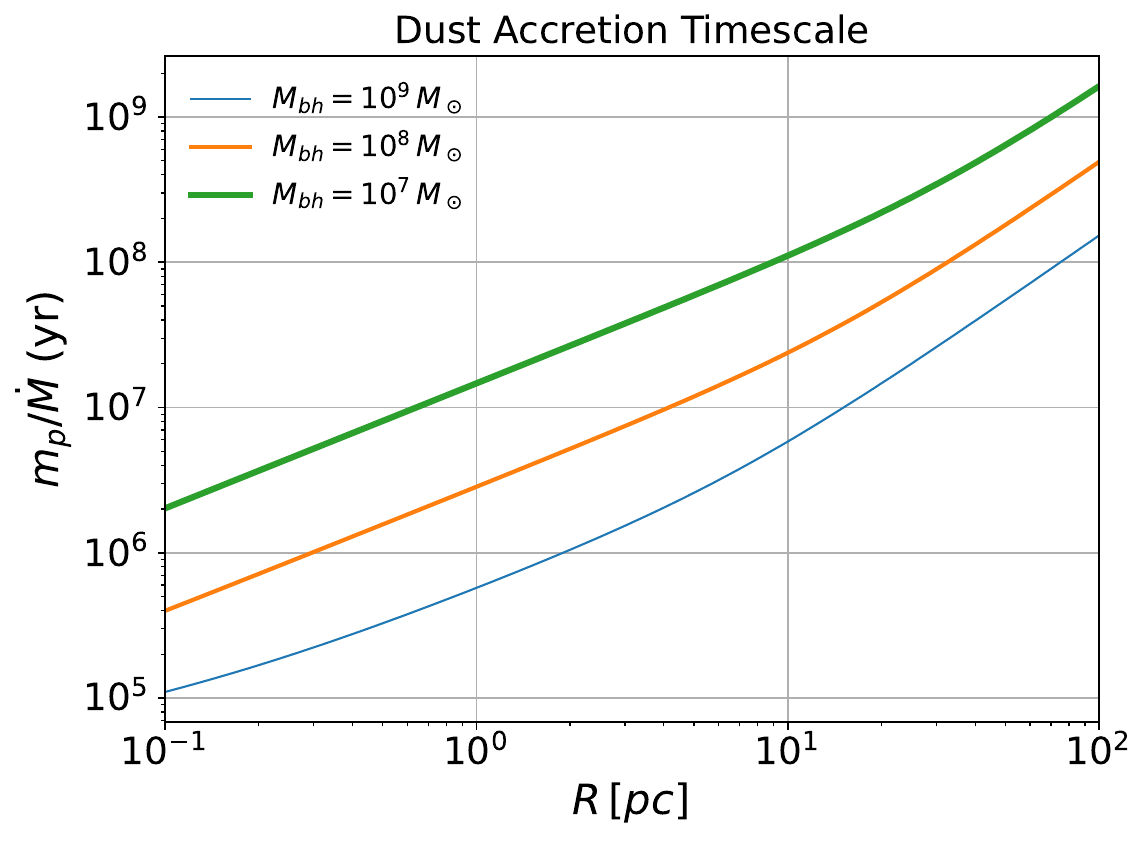} 
      \includegraphics[width=\columnwidth]{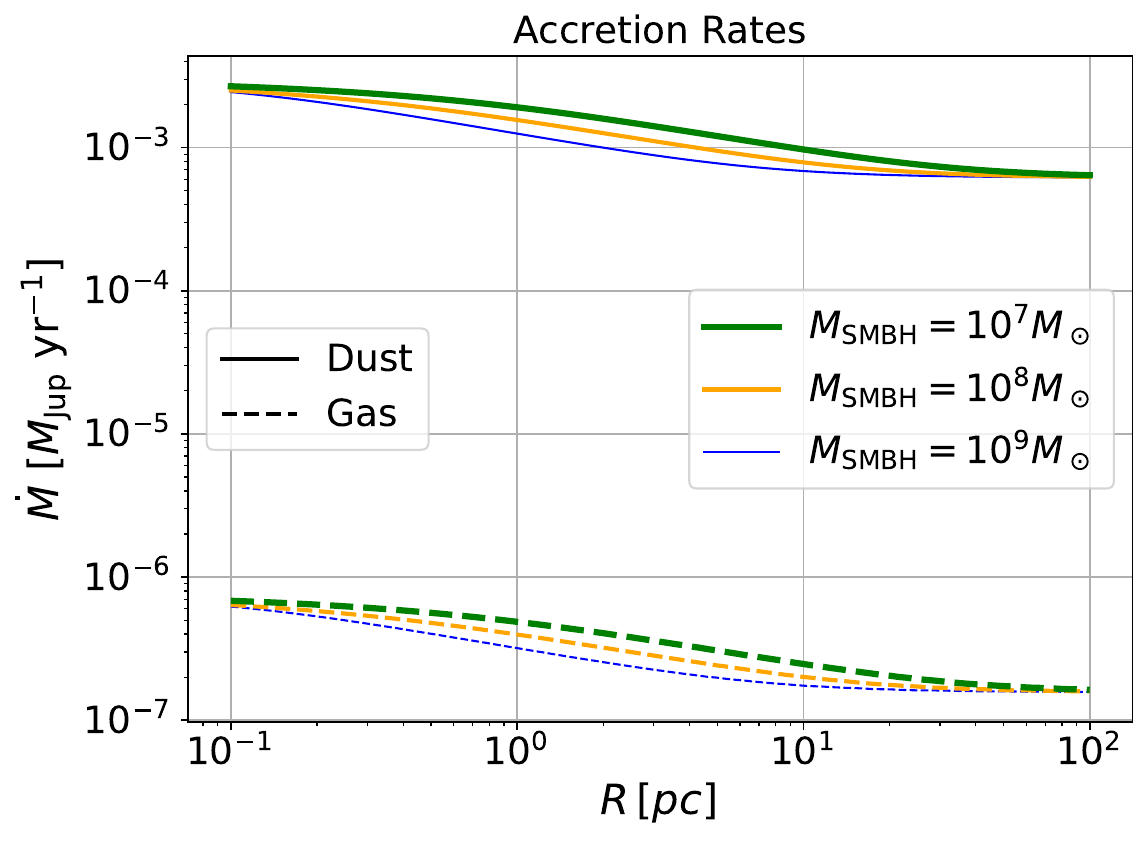}
      \includegraphics[width=\columnwidth]{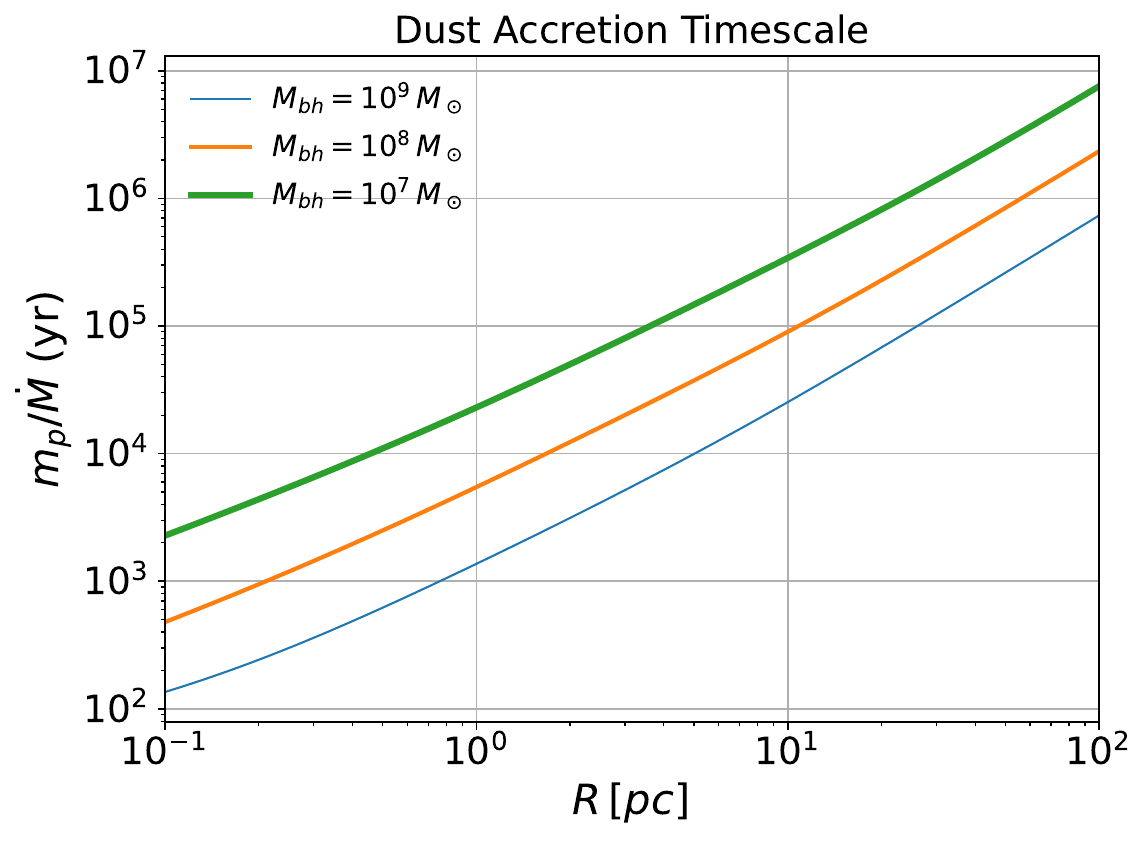}
    \caption{The e-folding growth time for the planets given the planetesimal mass at formation and pebble accretion rate. The top row shows $v_\mathrm{frag} = 1\, \mathrm{m\,s^{-1} }$ and the bottom row $v_\mathrm{frag} = 10\, \mathrm{m\,s^{-1} }$.}
    \label{fig:dustacc-timescale}
\end{figure*}

\begin{figure*}
    \centering
    \includegraphics[width=0.48\textwidth]
    {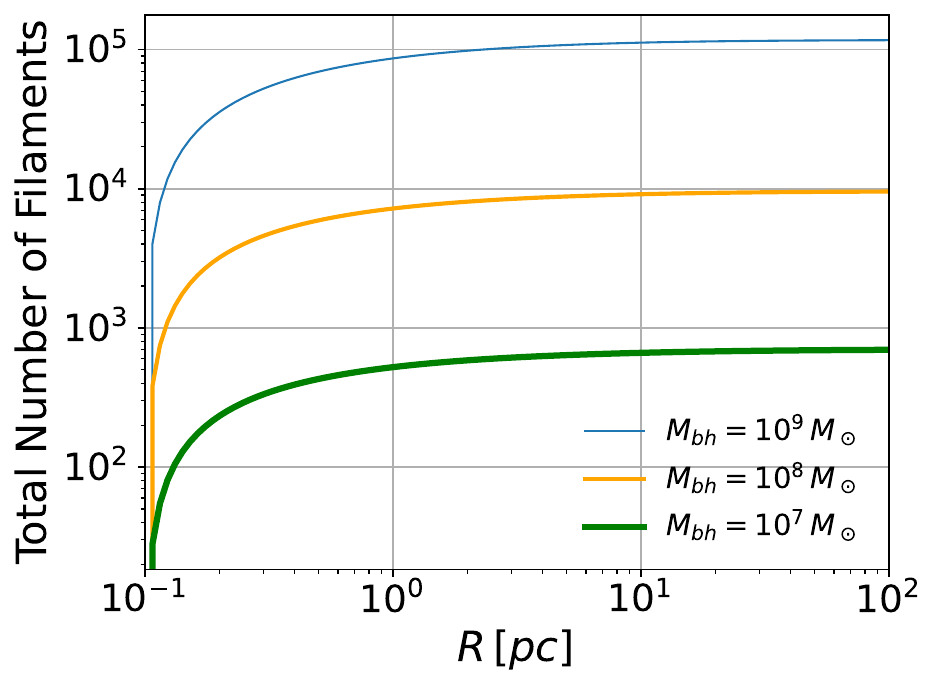}
    \includegraphics[width=0.5\textwidth]
    {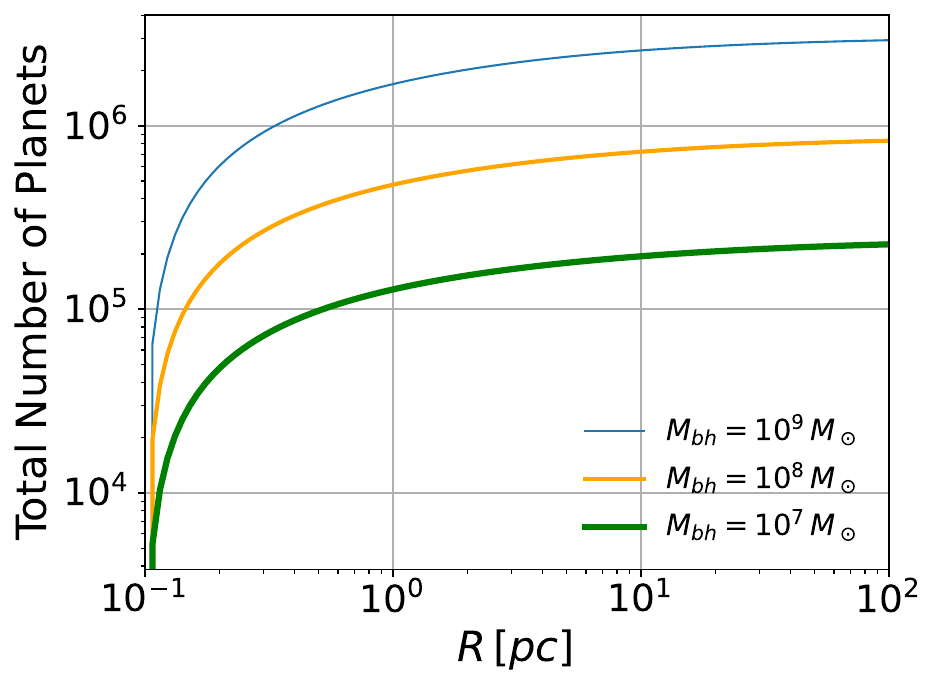}
    \caption{Cumulative sum over radius of number of filaments {\em (left)} and total number of planets {\em (right)}. }
    \label{fig:cumsum_totalplanets}
\end{figure*}

\subsection{{Fate of the seeds}}

\subsubsection{{Unusual objects: Jovian-mass and stellar-mass objects of pure dust}}

These planetary mass objects form out of the dust component of the AGN disk, so they have metallicity $Z=1$ (pure heavy elements) at zero age.  Yet, the mass of the objects seen in \fig{fig:streaminginstabilitymbhStSI1e-2vfrag1000}g and \fig{fig:mdot_hillbondiSt1e-2Vfrag100} are in the giant planet regime for the regions of the disk inward of 1 pc, and crossing the hydrogen burning limit for most of the parameter space beyond 10 pc---they are stellar mass rocky objects. These totally metal stars have been predicted before \citep{Hopkins14}, in the context of aerodynamical decoupling of grains in the turbulent ISM. That work concluded their formation was possible, but they would be exceedingly rare (1--$10^6$ in the Milky Way). Yet, in our model, they form the majority of the streaming instability output. While the objects formed in the Hill accretion regime, and perhaps also those formed in the Bondi regime, will increase their mass by pebble accretion, and may undergo substantial gas accretion to become bona fide stars, these stellar-mass rocky planets that are formed in the geometric regime would not accrete significantly, retaining their initial mass through the lifetime of the AGN. 

The geophysical structure and evolution of these objects is unknown, and probing it is beyond the scope of the current work. However, they are quite likely to have degenerate cores, since predominantly silicate objects have an electron to baryon ratio a factor of two lower than hydrogen, so they will reach degeneracy at half the total masses that a pure hydrogen object would, well under a Jupiter mass. 
Their outer layers are likely heated by
the radioactive decay of $^{26}$Al, $^{60}$Fe, and other short-lived radionuclides produced by massive evolved stars in the disk, so silicates would melt throughout, and the objects would have a magma ocean with an outgassed atmosphere. These would be degenerate lava drops orbiting in the AGN torus. We caution that the characteristic planetesimal mass is only the mass that becomes gravitationally bound, so it is technically a pebble cloud of negative energy, and further fragmentation could occur under the grid scale of the simulations, producing more objects of smaller mass \citep{PolakKlahr23}.

\subsubsection{Star formation}

Massive stars should grow fast and then die normally in a supernova. Immortal stars \citep{Cantiello+21} may only occur in narrow regions of AGN disk parameter space, if at all \citep{Fabj25}.
It is unknown how long AGN disks live. It is believed that the integrated activity in AGN in a given nucleus is $\mathcal{O}(100)$~Myr \citep{Soltan82}. However, it is likely that this activity is divided over multiple episodes, which may be either short ($<1$~Myr) or long ($>10$~Myr). Over long AGN timescales, gas accretion will allow for the formation of massive objects, potentially reaching the hydrogen burning limit and forming stars:  a core accretion channel for star formation. 

The onset of nuclear fusion would introduce stellar feedback mechanisms, such as stellar winds and ionizing radiation, which can ultimately suppress further accretion by dispersing the surrounding gas and even halting material infall onto the central forming object \citep{Jermyn22}. Thus, the lifecycle of planet and star formation in AGN disks could be self-regulating and limited by feedback processes once the accreted mass surpasses the hydrogen burning limit. Yet, some objects may accrete beyond this limit, and reach pebble isolation mass, terminating solid accretion, and then thermal mass
(which are of the same order of magnitude). At the thermal mass, the gravity of the object significantly affects surrounding gas dynamics and also overcomes the AGN disk pressure. The object's Hill radius becomes comparable to the disk's vertical scale height, so it opens a gap in the disk, creating pressure gradients (gap walls) that may further regulate accretion flows. For the massive torus in our model, this threshold is around $M_{\rm thermal} \sim 10^4\,M_\odot$. 
 
\subsubsection{Black hole formation}

Gas accretion allows newly formed gaseous protostars embedded within AGN disks to continue growing by accreting further gas from the dense, hot AGN disk. As these protostars evolve, sustained accretion can push their mass well beyond typical stellar values, altering their evolutionary pathways \citep{Cantiello+21}.

Once these protostars accumulate sufficient mass, they evolve toward the core-collapse phase at the end of their nuclear burning lifetimes of less than a few million years \citep{Fabj25},  well within longer AGN disk lifetimes. A diffuse scattered population of planets surrounding the AGN can help nucleate high metallicity clouds in the Broad Line region \citep{Hamann99,Nagao06}. The ultimate fate of such massive stars is strongly mass-dependent. For stars in the 30--300~$M_\odot$ range, core-collapse may lead to BH formation via fallback during or after a supernova explosion. For more massive stars ($>300\,M_{\odot}$), direct collapse to IMBHs may occur without a significant explosion \citep{RenzoSmith24}.

If accretion continues unabated and the protostar exceeds a mass threshold of around $300\,M_\odot$ prior to core collapse, current stellar evolution models \citep{Hosokawa09a, Hosokawa09b} suggest that the pair-instability mechanism \citep{Heger02}, which typically disrupts stars in the 140--260~$M_\odot$ range, may no longer operate effectively. In this regime, the core becomes gravitationally unstable \citep{Arnett89, Janka12}, and the energy from the supernova explosion may be insufficient to reverse the collapse. This leads to the formation of a massive BH through complete stellar collapse.

Such scenarios are a plausible pathway for the formation of IMBHs, particularly in the mass range $10^2$-–$10^5\,M_\odot$, which are otherwise difficult to explain through standard stellar evolution alone. The high-density environment of AGN disks, combined with sustained gas inflow and potentially episodic or chaotic accretion, provides an ideal setting for such massive object growth.

Additionally, these IMBHs may serve to feed SMBHs in galactic centers, especially if they form at high redshift and experience subsequent mergers or accretion-driven growth. This underscores the role of AGN disks not just in star formation, but also as active sites of BH assembly in the broader context of galaxy and structure formation in the universe.

\section{Discussion and Conclusions}
\label{sec:disscussion}
In this study, we explore the potential for planet and star formation in the outer regions of AGN accretion disks assuming a magnetically-supported disk model \citep{Hopkins24}, and low levels of turbulence. Our analysis focuses on regions beyond 0.01 pc, where the disk is cool and optically thin enough for dust survival and growth, which is a key condition for planet formation.

    
We find that the outer AGN disk environment can support dust coagulation and formation of planetesimals with masses exceeding that of Jupiter, up to and above the hydrogen burning limit, driven by the streaming instability. Despite the high masses, these objects are pure dust. We notice also that, unlike in the solar system, where the planetesimal masses only weakly scale with distance (as evidenced by the fact that asteroids and Kuiper-belt objects have similar dimensions), in the AGN case there is a strong radius dependence. This is understood given the scaling of the planet masses with disk aspect ratio $h^3$ in \eq{eq:mminLiu} and \eq{eq:mmin}. Whereas the aspect ratio in circumstellar disks is only slightly flared ($d\ln h / d\ln r \sim 2/7 \approx 0.25$, e.g.\ \citealt{ChiangGoldreich97}), an AGN torus is strongly flared, with a $d\ln h / d\ln r \sim 0.5$ dependency \citep[e.g.][and Fig~1f]{Moranchel-Basurto+21}. Cubing, we have $h^3 \propto r^{1.5}$, which matches the radial dependency seen in the planetary masses in Fig.~2g.

For a range of SMBH masses $10^7$--$10^9\,M_\odot$ at least, dust grains can grow by coagulation to sizes between millimeters and centimeters, with growth timescales well within the even short estimates of AGN disk lifetimes of 1--10~Myr.  By considering the mass of dust filaments produced in the nonlinear saturated state of the streaming instability, we estimate tens of millions of such objects could form, with their numbers increasing for higher SMBH masses and larger radial distances. Gas and pebble accretion allow accumulation of substantial further mass, depending on the radial location and central BH mass. We find that the pebble isolation mass is above the hydrogen burning limit, so pebble accretion in AGN should be halted by stellar feedback, when the planetary mass objects become stars.

The covering fraction of the distribution of clouds of gas and dust in the AGN torus may change over time \citep{Nenkova08}. The number of clouds $N_{\rm cl}$ in the AGN torus may limit the isolation mass and thermal mass available to accretors.
Dynamical interactions among this population of massive planetesimals within a mutual Hill sphere $R_{\rm Hill}$ (Eq.~\ref{eq:Hill})
should scatter planets and drive diffusion of this population both inwards and outwards. 
    
To provide observational support for our hypothesis, we can consider occultation and microlensing. These are two sides of the same process \citep{Agol02} and can be used to test for the presence of a numerous, scattered population of solid objects around an AGN. Such effects are more detectable against the small, central, X-ray emitting region of the AGN \citep{McKernan98,BekyKocsis13}. Thus, a search for eclipses and symmetric flaring in AGN lightcurves, particularly for smaller inner disks in shorter wavelengths such as UV and X-rays, should be a strong test of the population density of the scattered population of planets, stars, and BHs born in AGN tori. A diffuse scattered population of planets surrounding the AGN can help nucleate high metallicity clouds in the Broad Line region \citep{HamannFerland99,Nagao+06}.
    
The final masses of these objects could be calculated by integrating the dust and gas accretion rates concurrently; yet such final mass will depend strongly on the AGN lifetime, a quantity that is poorly constrained and varies widely. We thus simply divide the objects into those whose mass doubling time is lower than the AGN lifetime, and thus do not grow significantly, and the opposite case, the objects that do grow. When the mass doubling time is short compared to the AGN lifetime, dust and gas accretion may push some planetary-mass objects toward stellar mass. How stars grow in AGN is poorly constrained, but rapid accretion and growth to potentially very high mass is expected \citep{Cantiello+21,Jermyn22,Fabj25}.  Massive seed planets in the AGN disk can accrete enough material to exceed thermal and isolation masses of $\sim 10^4 M_\odot$, potentially transitioning into stars and eventually BHs.
Very massive stars $\mathcal{O}(100M_{\odot})$ are likely to last $<1$~Myr, triggering core collapse SNe, which should leave behind BHs. For accreted masses above $\sim 300\,M_\odot$, direct collapse into IMBHs becomes a viable outcome \citep{RenzoSmith24}, suggesting AGN disks as plausible birthplaces for such remnants. 
    
Dynamical interactions among the population should drive a mass segregation effect via equipartition, where the more massive population tends to sink inwards and the less massive component sinks outwards. Thus, IMBH and massive stars may sink inwards towards the inner disk. {Our model also predicts a significant population of stellar-mass objects composed of pure dust, for which little understanding of structure and evolution exists.}
    
If LIGO-VIRGO-KAGRA observes gravitational waves from a merging binary BH where one of the binary masses is $M_{1}>300\,M_{\odot}$, then either this is a direct formation BH, or one that has been built up through repeated hierarchical mergers. In the latter case, we might expect a high spin, but for the former, a potentially low natal spin. AGN tori are a plausible site for where such IMBHs might arise.  

In conclusion, AGN disks are favorable sites for the growth and formation of many astrophysically interesting objects from Jupiter-mass planets to stars, and stellar- or intermediate-mass BHs.
The outer regions, governed by dust dynamics, turbulence suppression, and efficient accretion mechanisms, appear to be a compelling physical analog to protostellar disks, albeit on vastly larger dynamical and thermal timescales. This work presents strong theoretical support for the existence of up to tens of million Jupiters-mass planets and a potential IMBH formation channel in AGN disks, directly bridging the fields of planet formation and BH growth.

\section*{Acknowledgments}

B. Mishra acknowledges support from POLONEZ BIS grant 2022/47/P/ST9/02315 co-funded by the National Science Center and the European Union Framework Programme for Research and Innovation Horizon 2020 under the Marie Sklodowska-Curie grant agreement no. 945339. WL acknowledges support from the NASA Theoretical and Computational Astrophysical Networks (TCAN) program via grant 80NSSC21K0497, the NASA Emerging Worlds program via grant 80NSSC22K1419, and NSF via grants AST-2007422 and AST-2511672. B. McKernan and KESF are supported by NSF grants AST-2206096 and AST-1831415 and Simons Foundation Grant 533845. M-MML is partly supported by NSF grant AST23-07950 and NASA grant 80NSSC25K7117. We acknowledge discussions with Anders Johansen and Sebastian Lorek on clarifying the planetesimal masses from streaming instability simulations.

\bibliography{./master.bib}{}
\bibliographystyle{aasjournal}
\end{document}